# Engineering heat transport across epitaxial lattice-mismatched van der Waals heterointerfaces

*Emigdio Chavez-Angel,[1] Polychronis Tsipas,[2] Peng Xiao,[1] Mohammad Taghi Ahmadi,[3] Abdalghani Daaoub,[3] Hatef Sadeghi,[3] Clivia M. Sotomayor Torres,[1,4] Athanasios Dimoulas[2] and Alexandros El Sachat[1,2]\**

[1] Catalan Institute of Nanoscience and Nanotechnology (ICN2), CSIC and BIST, Campus UAB, Bellaterra, 08193 Barcelona, Spain

[2] Institute of Nanoscience and Nanotechnology, National Center for Scientific Research "Demokritos," 15341 Agia Paraskevi, Athens, Greece

[3] School of Engineering, University of Warwick, Coventry CV4 7AL, United Kingdom

[4] ICREA, Passeig Lluis Companys 23, 08010 Barcelona, Spain

*Corresponding author: a.elsachat@inn.demokritos.gr

**Abstract.** Artificially engineered 2D materials offer unique physical properties for thermal management, surpassing naturally occurring materials. Here, using van der Waals epitaxy, we demonstrate the ability to engineer extremely insulating ultra-thin thermal metamaterials based on crystalline lattice-mismatched $Bi_2Se_3/MoSe_2$ superlattices and graphene/$PdSe_2$ heterostructures with exceptional thermal resistances (70–202 $m^2$K/GW) and ultralow cross-plane thermal conductivities (0.01–0.07 W/mK) at room temperature, comparable to those of amorphous materials. Experimental data obtained using frequency-domain thermoreflectance and low-frequency Raman spectroscopy, supported by tight-binding phonon calculations, reveal the impact of lattice mismatch, phonon-interface scattering, size effects, temperature and interface thermal resistance on cross-plane heat dissipation, uncovering different thermal transport regimes and the dominant role of long-wavelength phonons. Our findings provide essential insights into emerging synthesis and thermal characterization methods and valuable guidance for the development of large-area heteroepitaxial van der Waals films of dissimilar materials with tailored thermal transport characteristics.

The recent advent of van der Waals (vdW) heterostructures and superlattices (SLs) has opened new perspectives in nanoelectronics with ultra-high mobility and topological properties, in optics with high absorption and sensitivity, as well as in the field of heat transport engineering, with low scattering rates and highly anisotropic properties.[1-4] Specifically, thermodynamically stable misfit layer compounds, vdW heterostructures and SLs with tailored thermal transport and thermoelectric conversion properties have been recently proposed for thermal management applications.[4-8] Due to the periodic nature of SLs, new phonon modes and bandgaps can be formed due to the folding effect, which results in strong modifications of the phonon group velocity and thermal conductivity depending on the period thickness.[9] More interestingly, vdW SLs assembled from layers with structural lattice mismatch and weak vdW interactions could enable strong reduction of phonon transport along the *c*-axis while retaining their in-plane crystallinity.

Despite significant efforts toward this direction, many experimental studies have reported vdW stacks using top-down fabrication methods, such as exfoliation, which can only prepare small flakes on a micrometer scale.[8,10-12] Additionally, such flakes is most likely to have defects or contamination. Nevertheless, Vaziri et al.[12] and Sood et al.[8] have demonstrated high thermal isolation across a few micrometer size exfoliated Gr/MoSe$_2$/MoS$_2$/WSe$_2$ heterostructures and graphene/MoS$_2$ SLs, respectively. In addition, Kim et al. have achieved ultra-low cross-plane thermal conductivity at room temperature (~ 0.041 W/mK) using van der Waals films with random interlayer rotations, but exclusively in polycrystalline WS$_2$ films.[13] In contrast, bottom-up epitaxial techniques, such as molecular beam epitaxy (MBE), enable the fabrication of high-order vdW SLs at wafer scale with atomically smooth and abrupt periodic interfaces.[1,14-16] The ability to control atomic layer thickness and chemical composition also allows the precise designing of the transport properties of the SLs. Moreover, due to the weak vdW interactions between layers, vdW epitaxy offers great flexibility for integrating atomic layers of distinct materials such as metals, semiconductors, superconductors, or insulators, without considering lattice-matching requirements.[1,17]

Here, using wafer-scale heteroepitaxial growth we demonstrate superior cross-plane thermal insulation based on atomically-thin crystalline vdW layered materials. Specifically, we direct grow on different substrates high quality lattice-mismatched Bi$_2$Se$_3$/MoSe$_2$ SLs and graphene/PdSe$_2$ heterostructures of varying thickness that exhibit tailored thermal transport properties at the atomic-scale. Combining contactless characterization techniques, e.g., frequency-domain thermoreflectance (FDTR) and low-frequency Raman spectroscopy, we study the acoustic and thermal properties of these epitaxial films. We focus on unraveling the impact of vibrational mismatch, phonon-interface scattering, temperature and film thickness on cross-plane thermal transport and we estimate the effective cross-plane thermal conductivity and total thermal resistance of the films taking into account all the interfacial contributions in our multilayer structures. Phonon transport calculations support our experimental data and further reveal the impact of the thermal contact resistance on thermal conduction of ultra-thin layered materials and the presence of different thermal transport regimes in such atomically thin 2D films. Untill todate, only the in-plane thermal transport properties of $Bi_2Se_3$, $MoSe_2$ and $PdSe_2$ films have been investigated;[18-22] while such high cross-plane thermal insulating properties have been achieved only using either polycrystalline films consisting of a single material or exfoliated vdW stacks.

**Results**

We grew the epitaxial $Bi_2Se_3$ films, $Bi_2Se_3/MoSe_2$ SLs and graphene/$PdSe_2$ heterostructures on various single-crystal substrates such as strontium titanate (STO), sapphire and silicon carbide (SiC) by MBE, yielding 2D thin films with minimal disorder. A schematic representation of each vdW material is shown in Fig. 1a-c. The structural and chemical characterization of the samples were studied by X-ray Photoelectron Spectroscopy (XPS), low-frequency Raman spectroscopy, scanning tunnelling microscopy (STM) and high resolution scanning transmission electron microscopy (HR-STEM) measurements. Figure 1d displays HR-STEM images of different thickness $Bi_2Se_3/MoSe_2$ SLs that confirm the expected layering structures, which consist of vertically stacked $Bi_2Se_3$ and $MoSe_2$ sublayers with atomically sharp and contamination-free interfaces. Reflection high-energy electron diffraction (RHEED) patterns show that $MoSe_2$ and $Bi_2Se_3$ layers are repeatedly grown highly oriented on top of each other despite their large lattice mismatch (~20%) (see also Fig. S1 in Supporting Information (SI)). The RHEED patterns of one quintuple layer (QL) $Bi_2Se_3$ and $Bi_2Se_3/MoSe_2$ heterostructure (Fig. 1e-h) show the difference in the relative positions of the streaks, which reflects the large lattice mismatch between $Bi_2Se_3$ and $MoSe_2$ at room temperature.

In situ XPS data for the $Bi_2Se_3$ thin films and $Bi_2Se_3/MoSe_2$ heterostructures are presented in Fig. 1k-n. The binding energies of the Bi $4f_{7/2}$ and Se $3d_{5/2}$ core levels for 1 QL $Bi_2Se_3$ grown directly on STO were 158.00 eV and 53.71 eV, respectively, in good agreement with previous reported[23] on thin-film single crystal $Bi_2Se_3$. After 2 ML $MoSe_2$ growth, the Mo $3d_{5/2}$ and Se $3d_{5/2}$ peak positions at 228.54 eV and 54.25 eV, respectively, indicate Mo-Se bonds and agreed well with the $MoSe_2$ formation.[24] The Se 3d peak in Fig. 1n is deconvoluted in four peaks to consider two types of bonds, namely, Bi-Se and Mo-Se bonds, keeping the Se $3d_{5/2}$-$3d_{3/2}$ spin-orbit splitting fixed at 0.86 eV. The two distinct Se environments suggest $Bi_2Se_3$ and $MoSe_2$ formation rather than the formation of a mixed Bi-Mo-Se compound. The latter is reinforced by the fact that the Bi $4f_{7/2}$ peak position remains the same after $MoSe_2$ film growth (see Fig. S2 in SI). The XPS spectra from the graphene/$PdSe_2$ heterostructures are shown in the SI (Fig. S3).

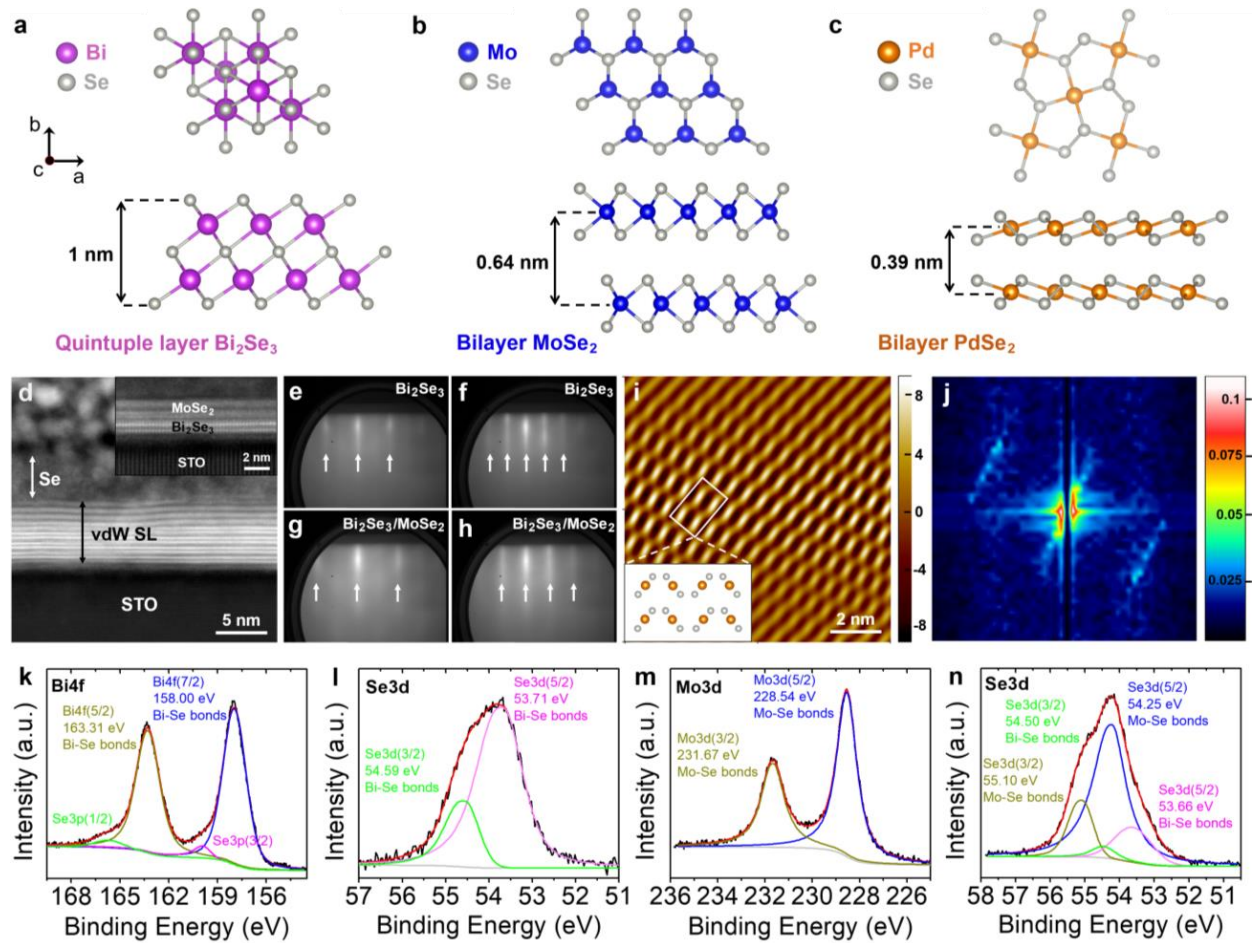
5

**Fig. 1. Structural and chemical characterization of epitaxial vdW films.** Top and side views of (a) $Bi_2Se_3$, (b) $MoSe_2$ and (c) $PdSe_2$ crystal structures. The purple, blue and orange spheres represent Bi, Mo and Pd atoms, respectively, while Se atoms are shown with grey spheres. 1 QL $Bi_2Se_3$ consists of five atoms per unit cell, such as Se-Bi-Se-Bi-Se, while each monolayer $MoSe_2$ consists of three atomic sublayers, in which Mo atoms are sandwiched between Se atoms. In $PdSe_2$, each Pd atom is connected with four Se atoms, and each Se atom is bonded with two Pd atoms and another Se atom. (d) Cross-sectional STEM images of the as-synthesized vdW $Bi_2Se_3/MoSe_2$ SLs with periods n = 3 and n = 1 (inset figure) on STO substrates. After the growth, a 4 nm thick Se capping layer was deposited in situ to protect the SL from oxidation. RHEED patterns of 1 QL $Bi_2Se_3$ and $Bi_2Se_3/MoSe_2$ SL with period n=1 along the (e), (g) [11–20] and (f), (h) [11–10] azimuths. (i) STM image of the as grown monolayer $PdSe_2$ on graphene and (j) the corresponding fast Fourier transform (FFT) image of the whole region. (k-n). XPS data of $Bi_2Se_3/MoSe_2$ SL grown on STO substrates.

Next, we systematically study the phonon properties of $Bi_2Se_3$ films, $Bi_2Se_3/MoSe_2$ SLs and graphene/$PdSe_2$ heterostructures using low-frequency Raman spectroscopy. Figure 2a shows the Raman spectra of $Bi_2Se_3$ films of varying thickness (from 1 to 20 QL), where all the out and in-plane Raman active optical modes are observed ($2E_g$ and $2A_{1g}$). Specifically, the $E^1_g$, $E^2_g$, $A^1_{1g}$ and $A^2_{1g}$ modes are detected at ~37 cm$^{-1}$, ~132 cm$^{-1}$, ~71 cm$^{-1}$ and ~173 cm$^{-1}$, respectively, in agreement with previous studies.[25,26] Both out-of-plane modes ($A^1_{1g}$, $A^2_{1g}$) show a pronounced red shift (about 2.7 cm$^{-1}$) as the thickness decreases; while the in-plane modes ($E^1_g$, $E^2_g$) are red-shifted with decreasing thickness of about 1.7 and 3.5 cm$^{-1}$, respectively (see also Fig. S11a). We note that the $A^1_{1g}$ is more sensitive with thickness because it reflects the out-of-plane vibrations of the Se and Bi atoms.[26] We also observe a broadening of the $E^2_g$ mode as the film thickness decreases (see Fig. 2a), in agreement with previous reports,[26] suggesting that the layer-to-layer stacking strongly affects the interlayer bonding. In SLs (see Fig. 2b) except of the Raman modes that correspond to $Bi_2Se_3$, we detect the $A_{1g}$ mode in the spectral range of 240.1–241.1 cm$^{-1}$, which confirms the existence of $MoSe_2$ layers.[27,28] In Figure 2c we also plot the ratios of the Raman intensities for the out-of-plane ($I_{A^1_{1g}}/I_{A^2_{1g}}$) and in-phase shear modes ($I_{E^1_g}/I_{E^2_g}$) of the adjacent layers versus thickness for the case of $Bi_2Se_3$ films and $Bi_2Se_3/MoSe_2$ SLs for direct comparison. We find that for the case of $Bi_2Se_3$ films, both $I_{A^1_{1g}}/I_{A^2_{1g}}$ and $I_{E^1_g}/I_{E^2_g}$ ratios decrease with decreasing thickness of less than 3 nm, in agreement with a previous study.[26] In the SLs, the values of the ratios $I_{A^1_{1g}}/I_{A^2_{1g}}$ and $I_{E^1_g}/I_{E^2_g}$ are similar to those calculated in $Bi_2Se_3$ films and remain almost constant with decreasing period.

In graphene/$PdSe_2$ heterostructures, we detected 6 main peaks in the high-frequency region (>130 cm$^{-1}$) that belong to $A^1_g$, $B^1_{1g}$, $A^2_g$, $B^2_{1g}$, $A^3_g$ and $B^3_{1g}$ phonon modes of $PdSe_2$ (Fig. 2d). These modes can be attributed to the intralayer vibrations of $PdSe_2$. The Raman peak positions of all phonon modes showed a red shift with increasing the number of layers, in agreement with previous CVD grown $PdSe_2$ films.[29,30] Interestingly, the intralayer vibration at 149.1 cm$^{-1}$, originating from the intralayer Se–Se bonds, exhibits sufficiently strong Raman intensity in 3, 5, and 7L $PdSe_2$, indicating its strong coupling to the electronic states.[29] Furthermore, we observe Raman-inactive modes in the frequency region between 50–130 cm$^{-1}$, which are activated due to the breakdown of translation symmetry in few layers, as has been recently shown.[30,31] The phonon modes detected in the frequency region between 100 to 300 cm$^{-1}$ are also consistent with a recent study where graphene/$PdSe_2$ heterostructures have been formed by exfoliation.[32] Finally, the two Raman peaks, $SiC_1$ and $SiC_2$ with 196.6 cm$^{-1}$ and 204.3 cm$^{-1}$, respectively, originated from the undoped SiC substrate.[33] We note that the mode $A_g^2$ is not well visible because it partially overlaps with the SiC$_2$ mode.

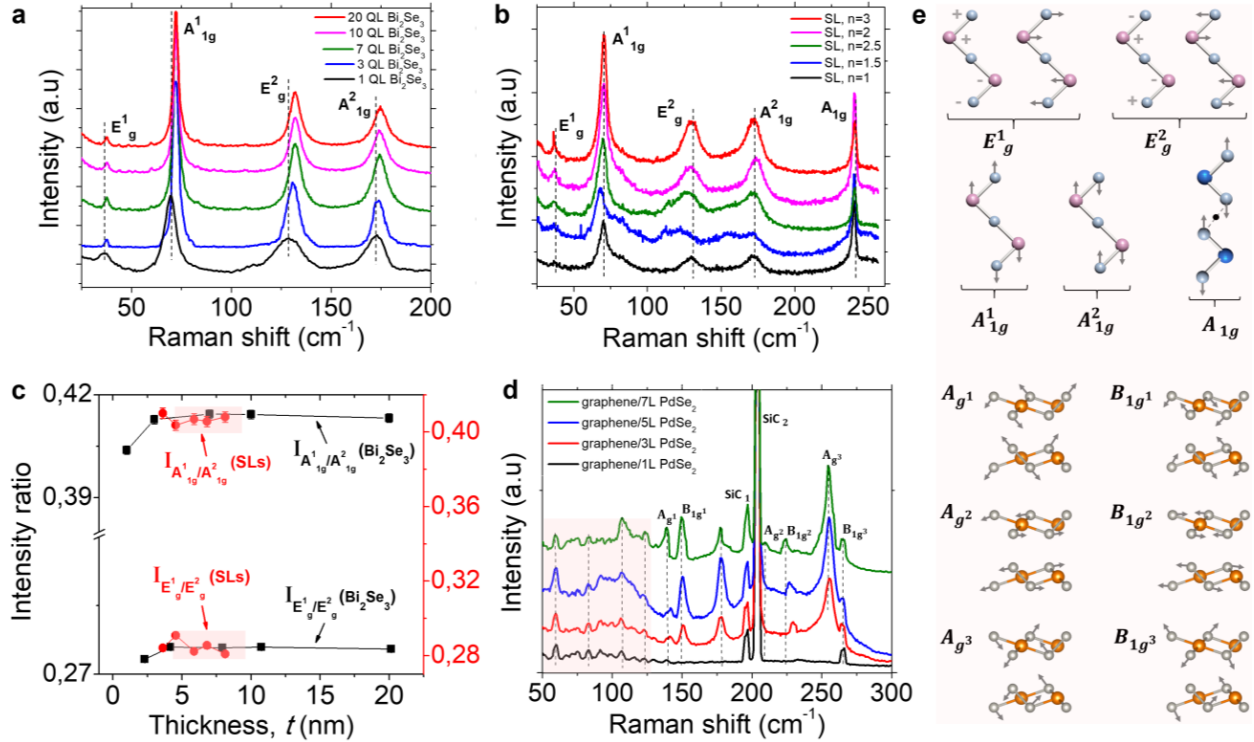

**Fig. 2. Low-frequency Raman spectroscopy in epitaxial vdW films.** Raman scattering spectra in (a) Bi$_2$Se$_3$ films of different thickness and (b) Bi$_2$Se$_3$/MoSe$_2$ SLs. (c) Ratios of the Raman intensities for the out-of-plane ($I_{A^1_{1g}}/I_{A^2_{1g}}$) and in-phase shear modes ($I_{E^1_g}/I_{E^2_g}$) versus film thickness for the case of Bi$_2$Se$_3$ films (black squares) and Bi$_2$Se$_3$/MoSe$_2$ SLs (red circles). (d) Low-frequency Raman scattering spectra in graphene/PdSe$_2$ heterostructures of different thicknesses. The red-shaded area shows the frequency region where the Raman-inactive modes are detected. (e) Atomic displacements (grey arrows) of all the Raman modes detected in Bi$_2$Se$_3$ films, Bi$_2$Se$_3$/MoSe$_2$ SLs and graphene/PdSe$_2$ heterostructures. The purple, blue and orange spheres represent Bi, Mo and Pd atoms, respectively, while Se atoms are shown with grey spheres.

The thermal measurements were performed using our custom-built frequency-domain thermoreflectance (FDTR) setup, which essentially combines simultaneous measurements of cross-plane thermal conductivity ($k_z$) and interface thermal conductance.[34,35] For the case of SLs, our multilayer structures consist of Au/SLs/substrate stacks (see Fig. 3a). For each SL, we obtained FDTR measurements and extracted the effective $k_z$ and the interface thermal resistances between Au (transducer)/SL ($R_1$) and SL/substrate ($R_2$) following a multilayer three-dimensional heat diffusion model.[36] The required material properties for this model are the thickness ($t$), the volumetric heat capacity ($C$), $R_1$, $R_2$ and $k_z$. The thermal conductivity and volumetric heat capacity of Au as well as the volumetric specific heat of $Bi_2Se_3$ and $MoSe_2$ were taken from the literature.[37,38] Therefore, the unknown parameters are the $k_z$, $R_1$ and $R_2$. To estimate these three parameters, first we quantified the sensitivity of the recorded phase signal to different parameters according to our multilayer geometry (see Fig. S4 in SI), following a similar methodology as reported elsewhere.[34-36]

Typical examples of the recorded phase signal and the corresponding best model fits in bilayer $MoSe_2$ and $Bi_2Se_3/MoSe_2$ SLs with periods 2, 2.5 and 3 are shown in Fig. 3b. To extract the $k_z$ of each SL from a single measurement, we followed the same approach that used in previous works,[34-36,39] and supported by our sensitivity analysis. First, we extract $k_z$ by fitting the experimental data in a low frequency range (20 kHz to 1 MHz), where the phase signal sensitivity to $R_1$, $R_2$, and $C$ of the films is negligible. Then, we fix $k_z$ and and fit the high frequency range (1−45 MHz) to estimate $R_1$ and $R_2$. The same procedure was followed to extract $k_z$, $R_1$ and $R_2$ for all the epitaxial films. The sensitivity analysis and FDTR data for the case of graphene/$PdSe_2$ are shown in Fig. S5 and Fig. S6, respectively. In Figure S9, we also show all the interface thermal resistance values, $R_{int} = R_1 + R_2$, where $R_1 = 1/G_1$ and $R_2 = 1/G_2$ extracted by the FDTR experiments and compare them with previous reports.

In SLs, we observe that $k_z$ slightly increases with increasing thickness with values between 0.059–0.07 W/mK (see Fig. 3c). However, in graphene/PdSe$_2$ heterostructures, we found a strong thickness-dependent $k_z$. Specifically, the $k_z$ increased by a factor of six with increasing the thickness of the top PdSe$_2$ layers from one to seven layers. To confirm the robustness of our approach to measuring the intrinsic $k_z$ of ultra-thin films, we performed FDTR measurements in Bi$_2$Se$_3$ films of different thicknesses in both STO and sapphire substrates. We found that in both substrates the $k_z$ of the Bi$_2$Se$_3$ films increases by a factor of five with increasing thickness from 1 to 20 QL. The excellent agreement in $k_z$ values is shown in Figure 3c.

The origin of the weak thickness dependence of the $k_z$ in SLs can be understood considering both interface-phonon scattering and size effects. By increasing the film thickness, thus the period of the SLs, thermal phonons (especially short-wavelength) are scattered by multiple Bi$_2$Se$_3$-MoSe$_2$ interfaces, thus reducing their contribution to cross-plane thermal transport. In contrast, the increase in the volume fraction of the constituents of the SLs with increasing thickness allows more long-wavelength phonons to propagate and contribute to $k_z$ until they are scattered at the sample boundaries. These opposite effects resulted in the suppressed thickness-dependent $k_z$ trend presented in Fig. 3c. Therefore, despite the influence of interface phonon scattering on $k_z$, the apparent increase of $k_z$ with increasing thickness in SLs indicates that in this thickness range finite size effects are still dominant. This is in agreement with previous studies that showed that the transmission of phonons across short-period AlN/GaN[40] and SiGe[41,42] SLs strongly depends on the phonon wavelength, suggesting that long-wavelength phonons are the dominant carriers of heat.

Molecular dynamics simulations have also shown that the $k_z$ of SLs with lattice-mismatched interfaces increases monotonically with period length.[43]

However, the absence or limited interfaces in pure $Bi_2Se_3$ and graphene/$PdSe_2$ allow the majority of the thermally excited phonons to contribute to cross-plane thermal transport, i.e., $k_z$ is limited mainly by finite size effects. This is reflected in the different rates of increase of $k_z$ observed in Fig. 3c. In particular, for the same thickness range, in $Bi_2Se_3$/$MoSe_2$ SLs, we found only a 28% of increase in $k_z$ with increasing thickness while in $Bi_2Se_3$ and graphene/$PdSe_2$ films of about 42% and 68%, respectively. We note that to study coherent and incoherent effects in SLs, the volume fraction of the constituents and total thickness of the films should remain constant while the thickness of each layer in a period must be adjusted to vary the interface density.[16] Therefore, in our SLs, the apparent linear scale of $k_z$ vs period is not sufficient to conclude if phonons travel coherently across the film thickness. For instance, previous calculations showed that despite the linear $k_z$ increase with thickness, lattice mismatch destroys the Bragg reflection conditions and phonons are diffusely scattered at interfaces, thus losing their coherency.[43]

To quantify the impact of cross-plane ballistic phonon transport on the total thermal resistance in our epitaxial films, we estimate the total thermal resistance per unit area, $R_{tot}$, which can be written as the sum of the combined interface thermal resistance, $R_{int} = R_1 + R_2$, and volumetric cross-plane thermal resistance, $R_{film} = t/k_z$.[34,44] From these calculations, we found that in the SLs $R_{tot}$ is linearly proportional to the total thickness (or number of periods, $n$), such that $R_{tot, n=3} > R_{tot, n=2,5} > R_{tot, n=2} > R_{tot, n=1,5} > R_{tot, n=1}$ (Fig. 3d). Furthermore, we observe that $R_{int}$ increases with increasing the period of the SLs (Fig. S9), in agreement with previous thermal resistance measurements in short-period SLs.[8,40] We note that the large lattice-mismatch between $Bi_2Se_3$ and $MoSe_2$ (~20%) most likely enhances the phonon interface scattering and further contributes to the increased values of $R_{tot}$. This is in agreement with previous studies that showed that lattice-mismatched interfaces exhibit reduced interface thermal conductance and phonon transmission due to the increased lattice disorder.[45,46]

However, in graphene/PdSe$_2$ heterostructures $R_{tot}$ remains constant with increasing the thickness of PdSe$_2$ from 1 to 7 layers. In fact, the different slopes in Fig. 3d suggest different thermal transport regimes. The similar values of $R_{tot}$ in graphene/PdSe$_2$ indicate strong ballistic thermal transport, as it is expected in a thin film with no internal scattering.[47] Similarly, cross-plane ballistic thermal transport has been found in graphene,[48] MoS$_2$[44] and PtSe$_2$[34] thin films. However, the increased values of the $R_{tot}$ with increasing thickness observed in SLs indicate additional phonon scattering (quasi-ballistic regime). We attribute this result to the scattering of high-frequency phonons at multiple Bi$_2$Se$_3$/MoSe$_2$ interfaces, which largely disrupt phonon ballistic transport. Therefore, interface roughness and lattice mismatch are effective in destroying the coherence of high-frequency phonons.

In Fig. 3d we show all our calculated $R_{tot}$ values and compare them with previous $R_{tot}$ measurements on different 2D materials and SLs, where similar interfacial contributions from bottom (2D material/substrate) and top (metal/2D material) interfaces were considered.[8,48] In Fig. S12 in SI, we also plot the volumetric cross-plane thermal resistance, $R_{film}$, as a function of thickness $t$, which further supports the previous discussion. Specifically, we observe that the $R_{film}$ is increasing by more than a factor of two as the thickness increases from 1.4 to 7.2 nm, which further suggests quasi-ballistic phonon transport and phonon diffusive scattering at interfaces. Notably, in ultra-thin graphene/PdSe$_2$ stacks, $R_{film}$ saturates at a finite value of about 54 m$^2$K/GW, .i.e., ballistic thermal transport. Similarly, Sood et al. using density functional theory (DFT) calculations, have estimated a constant cross-plane ballistic thermal resistance of ~10 m$^2$K/GW in MoS$_2$ films in the limit of 2−3 monolayers.[44] The significant contributions of both $R_{film}$ and $R_{int}$ components to the $R_{tot}$ in all the epitaxial films suggest that interfacial effects do not entirely govern cross-plane thermal transport.

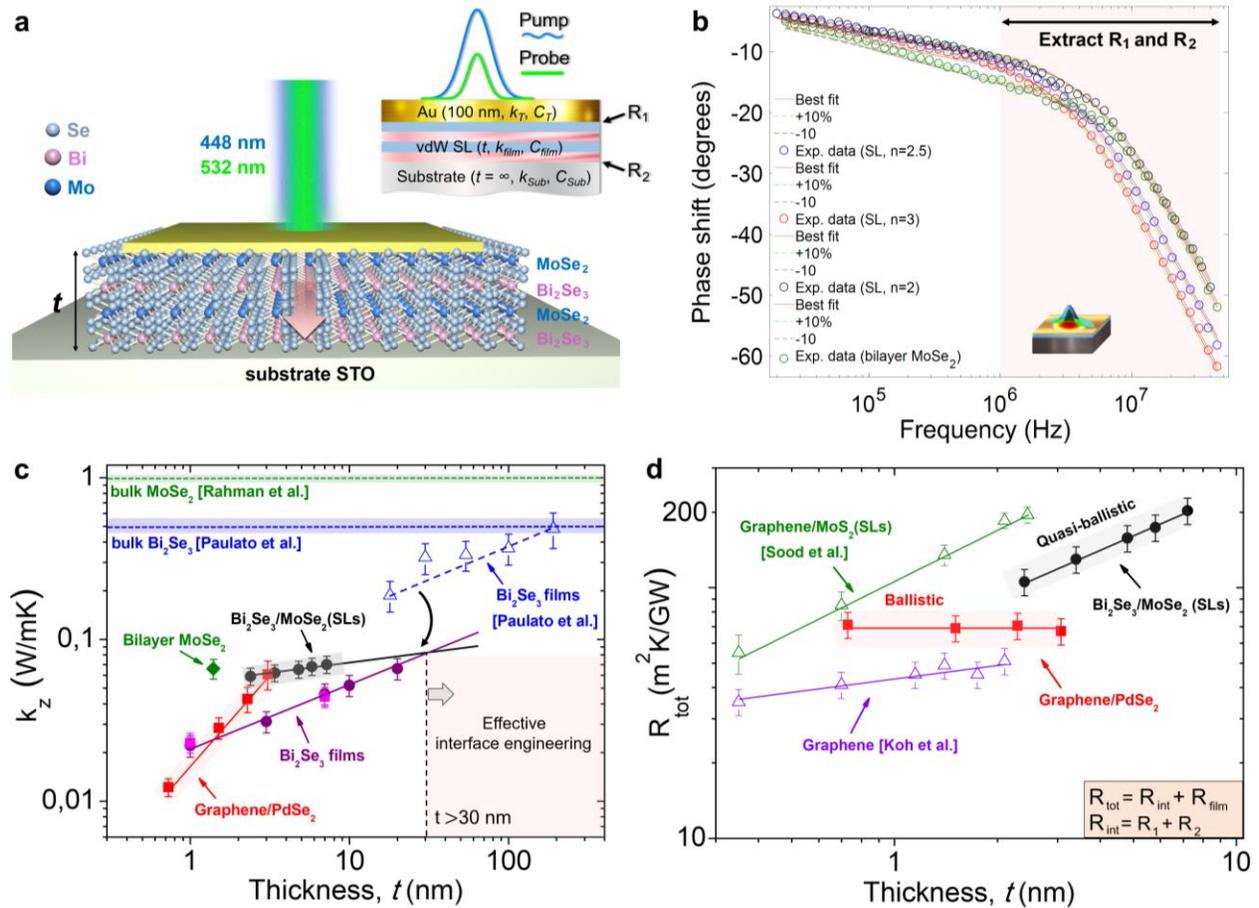

**Fig. 3. Thermal conductivity and interfacial heat transport measurements.** (a) Schematic illustrations of the FDTR technique and the multilayer system of the SLs. (b) Typical FDTR data measured in bilayer MoSe$_2$ and Bi$_2$Se$_3$/MoSe$_2$ SLs with periods $n$ = 2, 2.5, and 3 and the corresponding best model fits in the entire frequency range (20 kHz - 45 MHz). (c) Cross-plane thermal conductivity versus film thickness measured in bilayer MoSe$_2$ (green rhomb), Bi$_2$Se$_3$/MoSe$_2$ SLs (black circles), graphene/PdSe$_2$ heterostructures (red squares) and Bi$_2$Se$_3$ films on STO (purple circles) and sapphire (pink squares) substrates. The blue open triangles show

previously reported $k_z$ values measured in thicker $Bi_2Se_3$ films.[20] By extrapolating the measured $k_z$ trends, we find the minimum film thickness from which phonon-interface scattering starts to have a strong impact on cross-plane heat dissipation, i.e., $kz_{(SLs)} < kz_{(Bi2Se3)} < kz_{(MoSe2)}$. (d) Total thermal resistance, $R_{tot} = R_{int} + R_{film}$, of $Bi_2Se_3/MoSe_2$ SLs (black solid circles) and graphene/$PdSe_2$ hetrostructures (red solid squares) versus film thickness. The uncertainty of the estimated $R_{tot}$ was calculated on the basis of error propagation for the input parameters. Total thermal resistance measurements in Au/graphene/$SiO_2$[48] (purple open triangles) and Al/graphene-$MoS_2$ (SLs)/ $SiO_2$[8] (green open triangles) are included for comparison. The colored lines in (c) and (d) are guides for the eye.

To gain further insight into the influence of phonon-interface scattering on cross-plane thermal conductance, we constructed a tight-binding model[49] (see details in Methods) and calculated the transmission coefficients of phonons with different frequencies traversing through $MoSe_2$, $Bi_2Se_3$ and $Bi_2Se_3/MoSe_2$ heterostructures from one metallic electrode to the other (see Fig. 4a). The DFT calculations indicate that the Debye frequencies of $MoSe_2$ and $Bi_2Se_3$ are 24meV and 48meV, respectively.[50,51] We adjusted the parameters of our tight-binding model to obtain similar Debye frequencies for both materials (see Fig. S13 in SI). Furthermore, we took into account the lattice mismatch between the heterostructure layers by choosing a different coupling strength and coupling configurations between $MoSe_2$ and $Bi_2Se_3$ layers (see details in Figure S13 and Methods). Finally, to consider the effect of thermal contact resistance to electrodes on the overall thermal conductance, we considered two scenarios with weak and strong couplings to electrodes.

Figures 4b,c,d show the calculated $k_z$ of different thickness $Bi_2Se_3$, $MoSe_2$ and $Bi_2Se_3/MoSe_2$ heterostructures as a function of temperature. The $k_z$ increases with increasing temperature and starts to saturate at higher temperatures. Figure 4e shows the room temperature $k_z$ as a function of film thickness, which is in a good qualitative agreement with the experimental data when we consider a strong coupling between films and electrodes (see blue-shaded area). Specifically, the $k_z$ trends follow a similar order to that in the experiment, and unlike $Bi_2Se_3$, the $k_z$ in the SLs varies slowly with thickness. Interestingly, when we consider a weak coupling to electrodes, the calculated absolute $k_z$ values are in much better quantitative agreement with the experiments (see red-shaded area). This suggests that the total interface thermal resistance, $R_{int}$, is increased when the films are weakly coupled with the top and bottom electrodes, confirming the relative high $R_{int}$ values obtained from FDTR experiments (see Fig. S9). These results highlight the role of thermal contact resistance in cross-plane thermal conductivity of ultra-thin vdW layered materials.

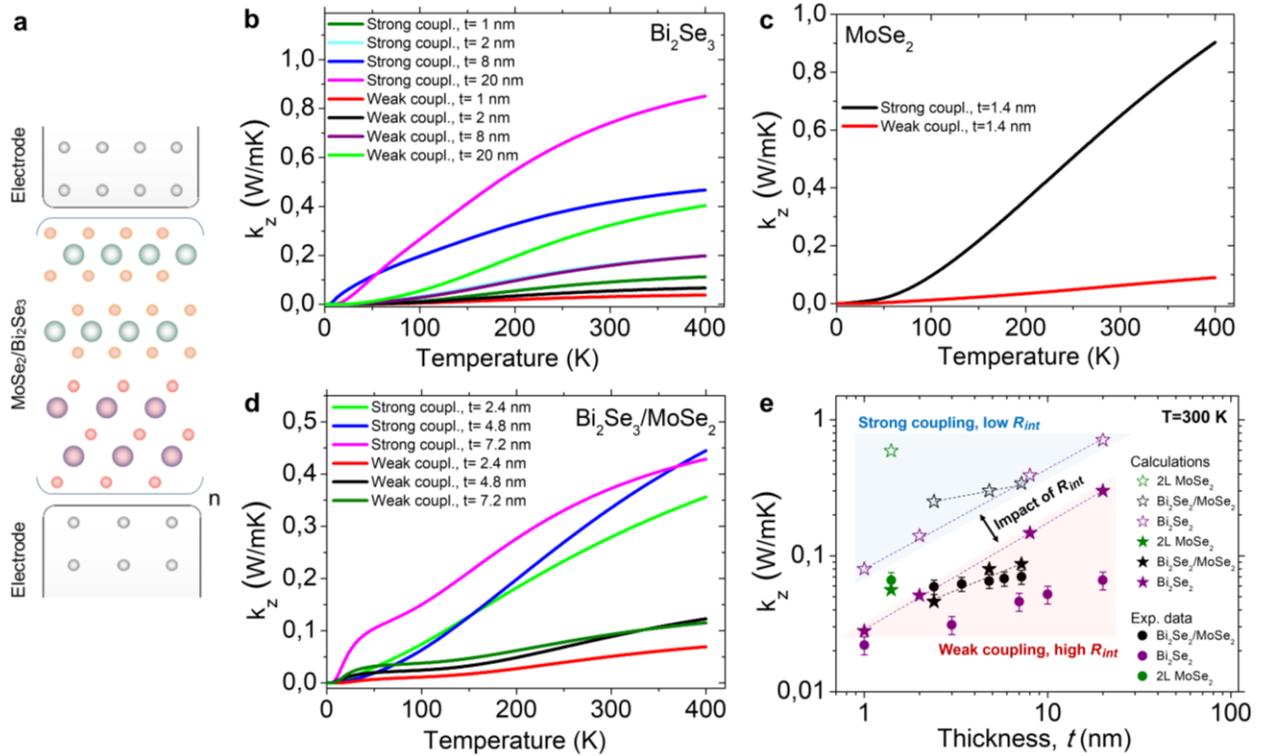

**Fig. 4. Tight-binding phonon calculations.** (a) Schematic structure of a $Bi_2Se_3/MoSe_2$ vdW layered structure between electrodes. The calculated cross-plane thermal conductivity of (b) $Bi_2Se_3$, (c) bilayer $MoSe_2$ and (d) $Bi_2Se_3/MoSe_2$ heterostructures of different thickness as a function of temperature (0-400 K) considering weak and strong coupling of the vdW films with the electrodes. (e) The calculated room temperature cross-plane thermal conductivity as a function of thickness for $MoSe_2$, $Bi_2Se_3$ and $Bi_2Se_3/MoSe_2$ heterostructures with weak (solid star symbols) and strong (open star symbols) coupling with the electrodes. The experimental $k_z$ values for $MoSe_2$, $Bi_2Se_3$ and $Bi_2Se_3/MoSe_2$ are shown with green, purple and black solid circles in (e) for direct comparison with the calculations. The difference in the calculated values indicated in blue and red shaded areas in (e) show the importance of quantifying interfacial thermal transport in vdW films bonded to a substrate.

In summary, we have successfully developed ultra-thin epitaxial vdW films that can act as highly insulating thermal metamaterials, comprising of dissimilar atomically thin layers of 2D semiconductors ($MoSe_2$ and $PdSe_2$), the 3D topological insulator $Bi_2Se_3$, and monolayer graphene. Our study reveals that short period crystalline $Bi_2Se_3/MoSe_2$ superlattices can be used to achieve a superior thermal resistance up to 202 $m^2K/GW$ and ultralow effective cross-plane thermal conductivity at room temperature down to 0.059 W/mK. We attribute this result to the interface roughness and large lattice mismatch between the constituent layers of the superlattices that boosts phonon-interface scattering and suppress cross-plane heat dissipation. Conversely, graphene/$PdSe_2$ heterostructures exhibit a strong thickness-dependent effective cross-plane thermal conductivity and a constant total thermal resistance of about 70 $m^2K/GW$ due to ballistic thermal transport. Given the sub-3 nm thickness of these heterostructures, their effective thermal conductivities at room temperature are estimated between 0.012 and 0.06 W/mK. Our phonon transport calculations align well with the experimental results and further shed light on the impact of interfacial thermal resistances between ultra-thin heterointerfaces and top and bottom metallic contacts on cross-plane heat dissipation.

Importantly, this work has yielded significant advancements in the epitaxial growth of high-quality heterogeneous interfaces over large areas, as well as in the quantitative understanding of interfacial thermal transport across atomically-thin vdW films on various substrates. The implications of these findings are extensive for the design of heat-sensitive electronic components and 2D electronic devices, such as 2D transistors and microchips, that can operate without thermal limitations e.g., overheating. The ability to obstruct heat dissipation in the vertical direction while maintaining in-plane crystallinity in wafer-scale engineered vdW stacks not only facilitates thermal management applications but also enhances their suitability as active materials in thermoelectric devices, thereby increasing their thermoelectric efficiency. Combining with bandgap engineering strategies, the followed phonon engineering approach could provide a promising route to realize a wide variety of functional semiconductor heterojunctions and superlattices for nanoscale electronic and thermoelectric devices.

## Methods

**MBE growth and in-situ characterization (XPS, STM)**

The experiments were carried out in an ultrahigh vacuum MBE system equipped with RHEED, X-ray photoelectron spectrometer and scanning tunneling microscope. High purity metals Mo and Pd were evaporated from an e-gun evaporator whereas Se and Bi were evaporated from effusion cells. The $Bi_2Se_3$ and $MoSe_2$ films were grown on STO (111) and sapphire substrates at ~300 °C under Se-rich conditions with a Se/Mo (Bi, Pd) flux ratio of about 20. The $PdSe_2$ films were grown on 4H-SiC (0001)/monolayer graphene substrates at ~240 °C under Se-rich conditions. The Mo, Bi and Pd growth rates were kept constant at ~0.1Å/sec. In-situ XPS measurements were performed with excitation by Mg Kα radiation (1253.6 eV) using a SPECS XR50 source. After $PdSe_2$ growth, the samples were transferred to the STM chamber without breaking the vacuum for in-situ STM characterization. STM images were obtained at room temperature using a Pt/Ir tip. The scanning conditions were V=0.4 Volt and I=400 pA.

**Raman spectroscopy**

The Raman spectra were recorded using a customized setup using a Monovista Raman spectrometer manufactured by Princeton and ensembled by S&I GmbH. It was used in a single grating mode (2400 lines) with a spectral resolution better than 0.4 cm$^{-1}$. The laser line was rejected using a Bragg filter around ±5 cm$^{-1}$. The samples were placed in an automatic xyz stage. Then, a green diode laser (λ = 532 nm) was focused on the sample using a 100× microscope objective. The power of the laser was kept as low as possible (<100 μW) to avoid any possible effect of self-heating.

**Frequency domain thermoreflectance**

FDTR is an ultra-fast laser based pump-probe technique which can measure thermal properties of bulk, thin films and nanostructured materials. The experimental setup is based on two lasers operating at 488 nm (pump) and 532 nm (probe). The pump beam is modulated in a wide frequency range (20 kHz–45 MHz), generating a periodic heat flux with a Gaussian spatial distribution on the 2D sample surface. The reflectivity of the sample, which is probed at the same position by the probed laser, changes as a function of the temperature and displays a phase lag compared to the pump signal. The phase response of the reflected probe beam to the thermal wave was recorded using a lock-in amplifier while the pump signal was used as a reference. We also used a 50× objective to repeat the FDTR measurements and obtained consistent results. The measurements were performed under ambient conditions at room temperature (∼22 °C). Before FDTR measurements, 100 nm Au films were deposited on top of the MBE-grown samples and used as transducers.

**Computational Method**

To model thermal conductance due to phonons, we first construct a tight-binding dynamical matrix of the MoSe$_2$ and Bi$_2$Se$_3$ lattices using the parameters shown in table S1 of the SI. The choice of these parameters is informed by comparing our TB Debye frequency with the Debye frequency computed using density functional theory.[50,51] We then construct the dynamical matrix of junctions including the layered materials between electrodes. Following the method described in a previous report,[49] the phonon transmission $T_p(\omega)$ then can be calculated from the relation $T_p(\omega) = Trace(\Gamma_L^p(\omega)G_p^R(\omega)\Gamma_R^p(\omega)G_p^{R\dagger}(\omega))$ where $\Gamma_{L,R}^p(\omega) = i(\sum_{L,R}^p(\omega) - \sum_{L,R}^p{}^\dagger(\omega))$ describes the level broadening due to the coupling to the left $L$ and right $R$ electrodes, $\sum_{L,R}^p(\omega)$ are the retarded self-frequencies associated with this coupling and $G_p^R = (\omega^2 I - D - \sum_L^p - \sum_R^p)^{-1}$ is the retarded Green's function, where $D$ and $I$ are dynamical and the unit matrices, respectively. The phonon thermal conductance $\kappa_p$ at temperature $T$ is then calculated from $\kappa_p(T) = (2\pi)^{-1}\int_0^\infty \hbar\omega T_p(\omega)(\partial f_{BE}(\omega,T)/\partial T)d\omega$ where $f_{BE}(\omega,T) = (e^{\hbar\omega/k_BT} - 1)^{-1}$ is the Bose Einstein distribution function, $\hbar$ is reduced Planck's constant and $k_B$ is Boltzmann's constant.

## Data availability

The data that support the findings of this study are available from the corresponding author upon request.

**Acknowledgements**

A.E. acknowledges funding from the EU-H2020 research and innovation program under the Marie Sklodowska Curie Individual Fellowship THERMIC (Grant No. 101029727). H.S. acknowledges the UKRI for Future Leaders Fellowship numbers MR/S015329/2 and MR/X015181/1. E.C.A and C.M.S.T. acknowledge support from the project LEIT funded by the European Research Council,

H2020 Grant Agreement No. 885689. ICN2 is supported by the Severo Ochoa program from the Spanish Research Agency (AEI, grant no. SEV-2017-0706) and by the CERCA Programme/Generalitat de Catalunya.

## Author contributions

A.E. conceived and supervised the research work. P.T. and A.D. fabricated the MBE samples, performed the XPS measurements and provided inputs in the structural analysis. A.E., E.C.-A. and P.X. performed the Raman and TEM measurements and analysis. A.E and E.C.-A. built the FDTR setup and performed the thermal measurements and data analysis. M.T.A, A.D. and H.S. performed the thermal conductance calculations and provided support to the theoretical analysis. All authors reviewed and edited the manuscript and have given approval to the final version of the manuscript. The manuscript was written by A.E. with inputs from E.C.-A., P.T., H.S. and C.M.S.T.

## Competing interests

The authors declare no competing interests

## Additional Information

**Supplementary Information**: RHEED patterns; XPS spectra; FDTR sensitivity analysis; FDTR experimental data; Raman experimental data; Tight-binding phonon calculations;

**Correspondence** and requests for materials should be addressed to Alexandros El Sachat.

# Supporting Information:

# Engineering heat transport across epitaxial lattice-mismatched van der Waals heterointerfaces

*Emigdio Chavez-Angel,[1] Polychronis Tsipas,[2] Peng Xiao,[1] Mohammad Taghi Ahmadi,[3] Abdalghani Daaoub,[3] Hatef Sadeghi,[3] Clivia M. Sotomayor Torres,[1,4] Athanasios Dimoulas[2] and Alexandros El Sachat[1,2]\**

[1] Catalan Institute of Nanoscience and Nanotechnology (ICN2), CSIC and BIST, Campus UAB, Bellaterra, 08193 Barcelona, Spain

[2] Institute of Nanoscience and Nanotechnology, National Center for Scientific Research "Demokritos", 15341 Agia Paraskevi, Athens, Greece

[3] School of Engineering, University of Warwick, CV4 7AL Coventry, United Kingdom

[4] ICREA, Passeig Lluis Companys 23, 08010 Barcelona, Spain

*Corresponding author: a.elsachat@inn.demokritos.gr

# 1. RHEED patterns and XPS data

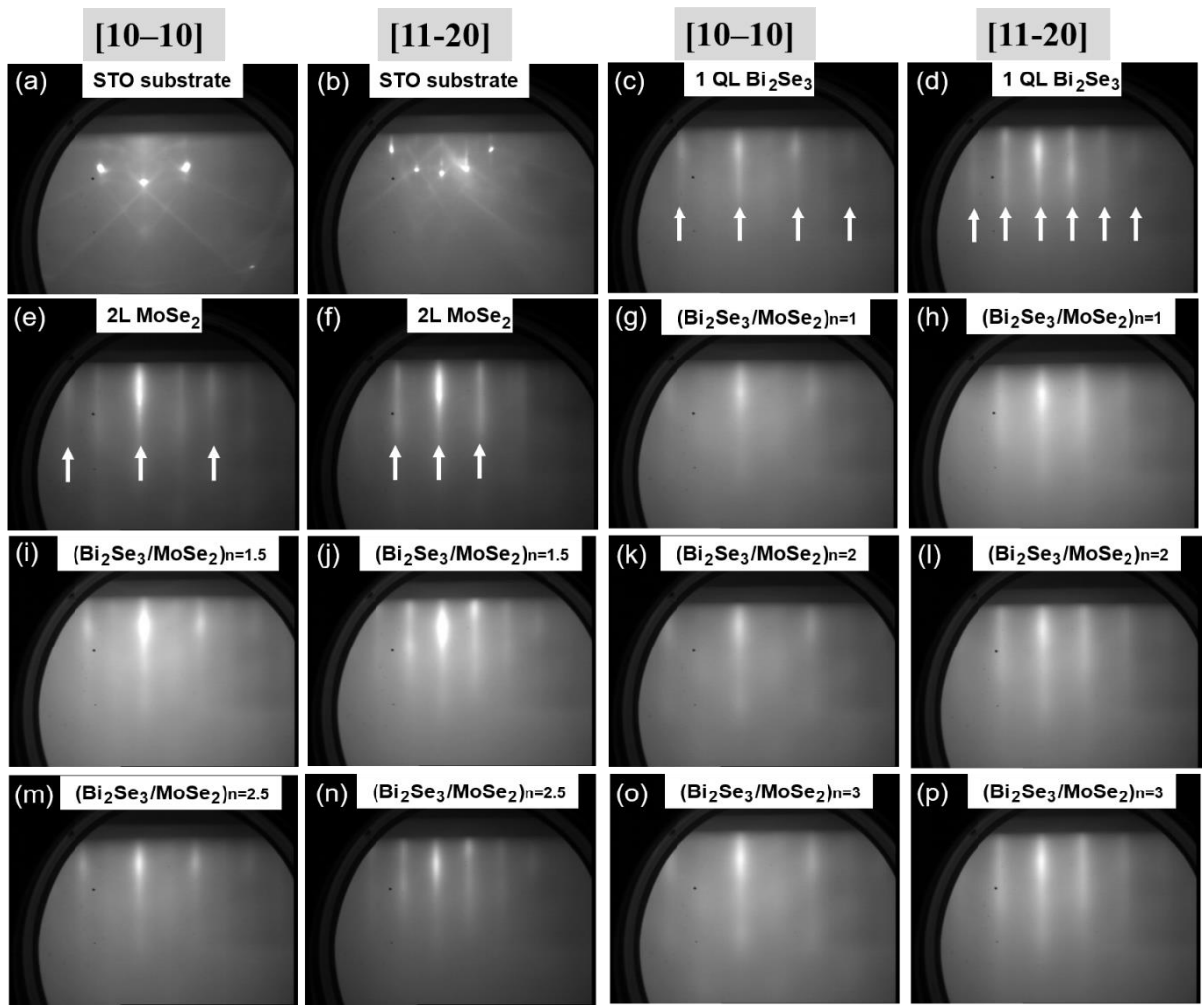

**Fig. S1.** RHEED patterns of all the $Bi_2Se_3/MoSe_2$ films along the [10–10] and [11-20] azimuths. The white arrows in (c), (d) and (e), (f) show $Bi_2Se_3$ and $MoSe_2$ streaks, respectively.

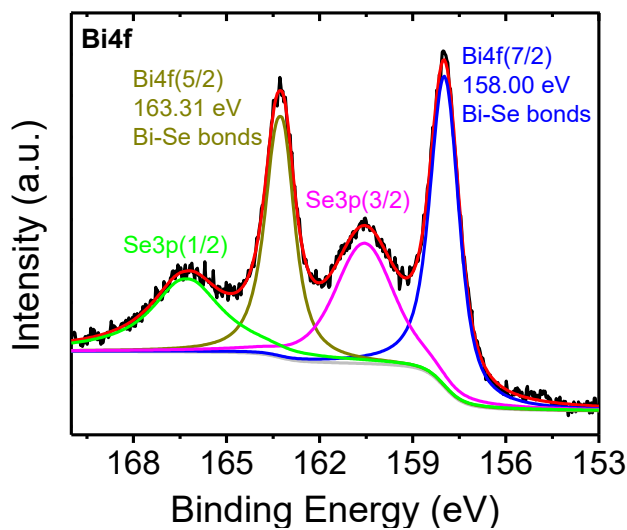

**Fig. S2.** XPS spectra in $Bi_2Se_3$/$MoSe_2$ heterostructure grown on STO substrate, showing that the Bi 4f (7/2) peak position remains unaffected after $MoSe_2$ film growth.

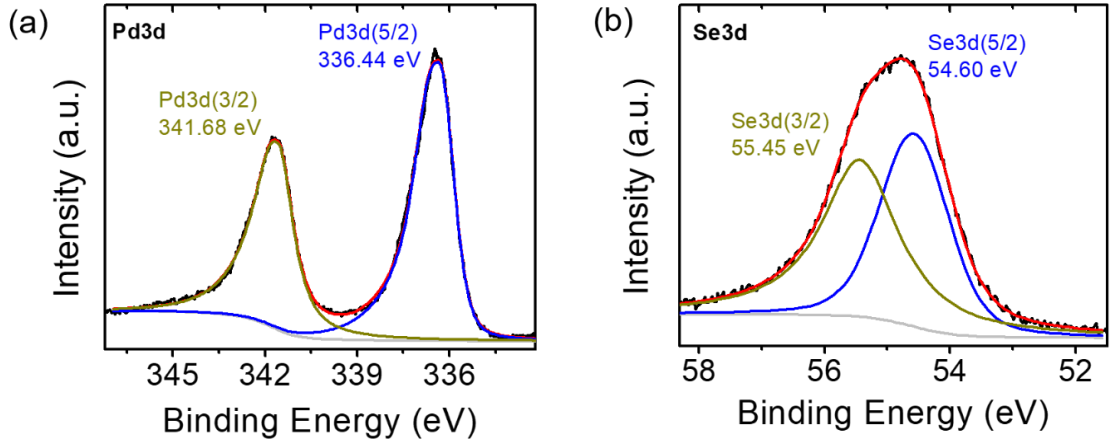

**Fig. S3.** (a) XPS data showing the Pd 3d (5/2) (336.44 eV) and Pd 3d (3/2) (341.68) core levels and Se 3d electrons. The Se 3d (5/2) (blue) and Se 3d (3/2) (yellow) peaks are overlapped and shown as one peak (black). No reaction with the substrate is observed and the positions and line shapes of the Pd 3d and Se 3d peaks indicate Pd–Se bonding.

## 2. FDTR sensitivity analysis and experimental data

For the sensitive analysis, we used a polynomial fit of the experimental thickness dependence of the cross-plane thermal conductivity ($k_z$). This fit allows us to have a continuous function to plot the sensitivity as a function of excitation frequency (20 kHz–45 MHz) and sample thickness. From Fig. S4a,b and Fig. S5ab we find that the sensitivity of the phase signal to $G_{12}$ =1/$R_1$ and $G_{23}$ =1/$R_2$ is relatively low and mainly at high frequencies. However, the sensitivity of the phase signal to $k_z$ is high in all the frequency range (Fig. S4d and Fig. S5d). In our measurements, the error bars of the $k_z$, were estimated by the standard deviation of several measurements, including the numerical errors from the fits (see Fig. S6-S8).

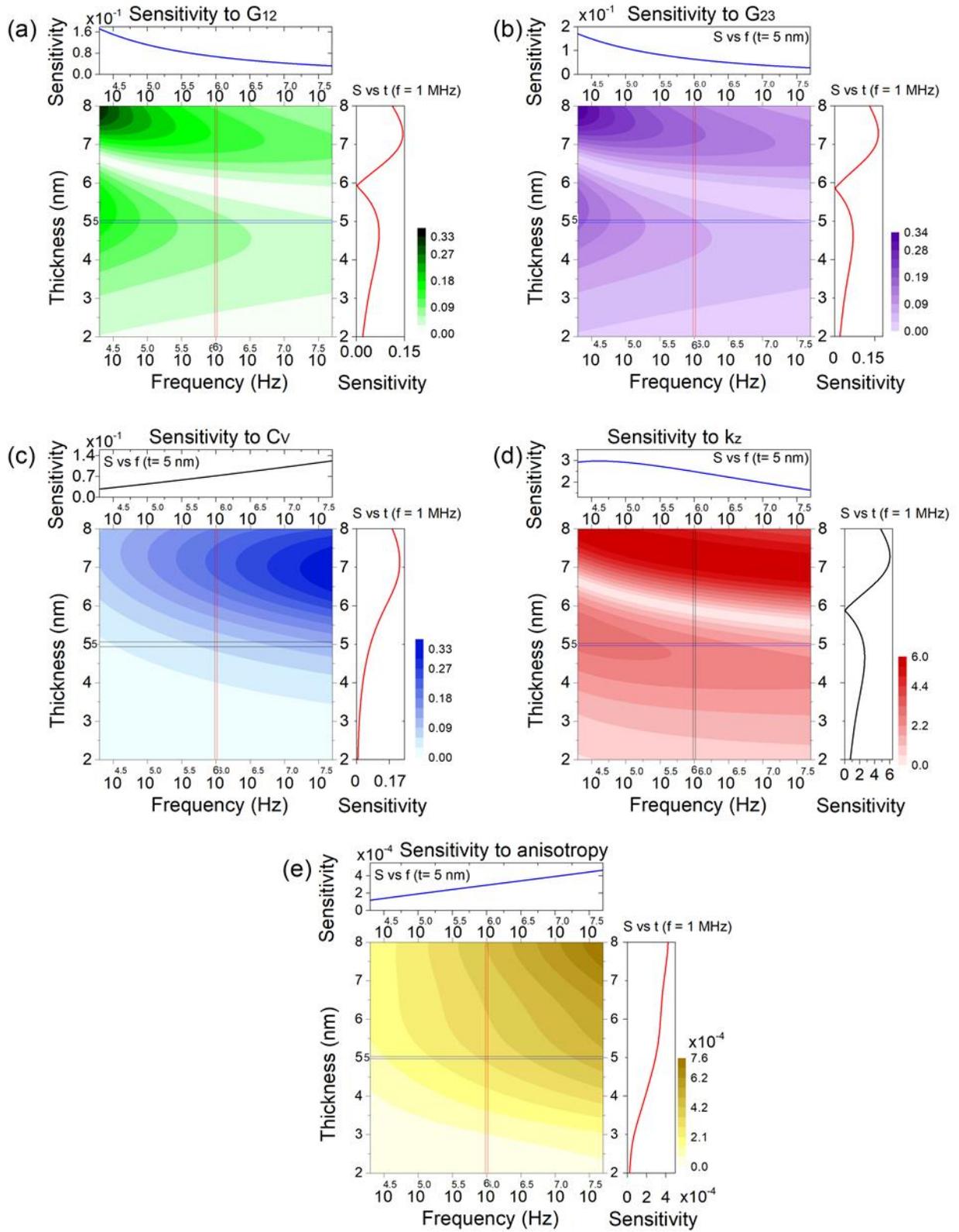

**Fig. S4.** Calculated phase sensitivity ($-V_{in}/V_{out}$) to different parameters, (a) $G_{12}=1/R_1$ (b) $G_{23}=1/R_2$ (c) $C_v$, (d) $k_z$, and (e) anisotropy as a function of thickness and modulation frequency for the case of Au/SLs/STO stacks.

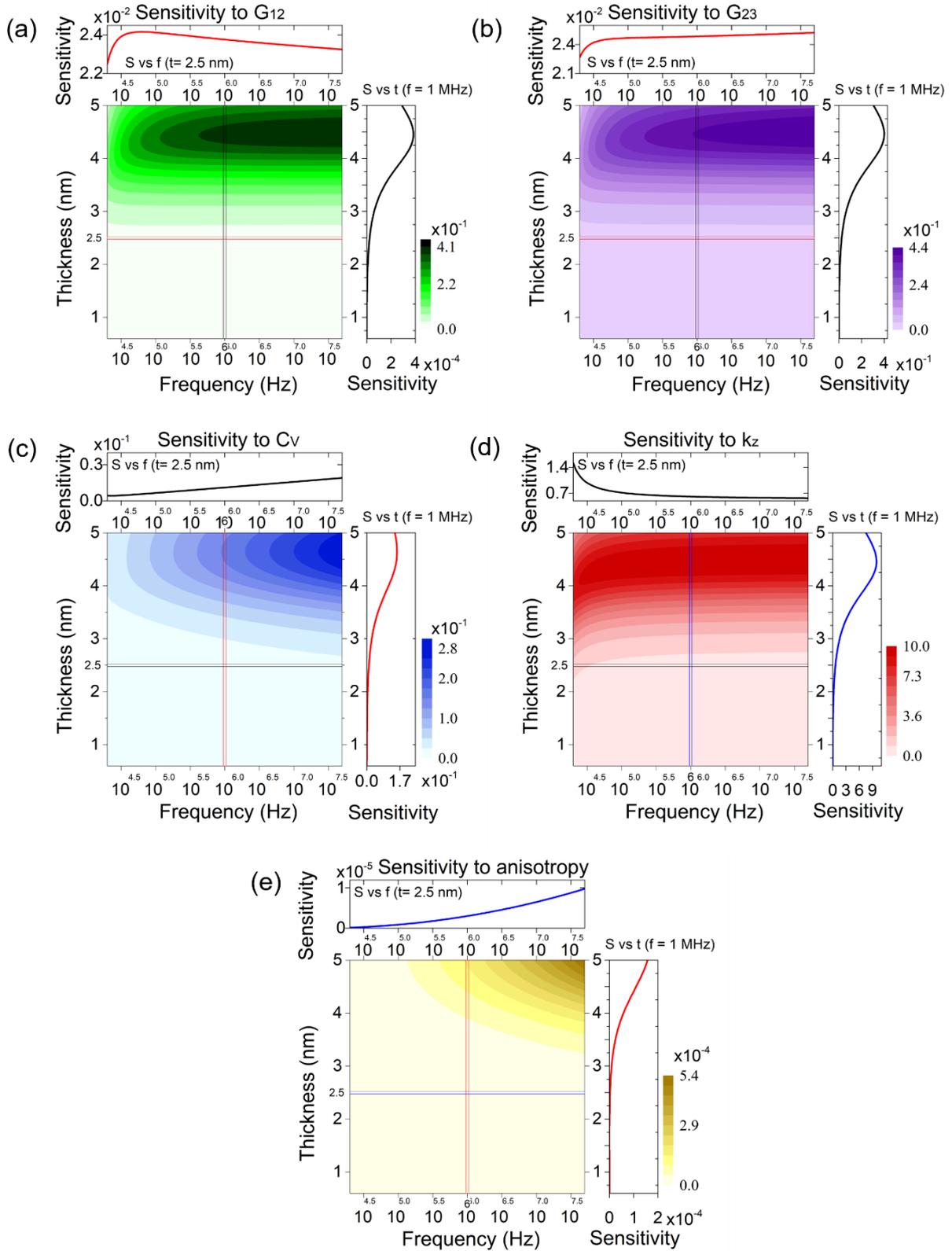

**Fig. S5.** Calculated phase sensitivity ($-V_{in}/V_{out}$) to different parameters, (a) $G_{12}$, (b) $G_{23}$, (c) $C_v$, (d) $k_z$, and (e) anisotropy as a function of thickness and modulation frequency for the case of Au/PdSe$_2$/graphene/SiC stacks.

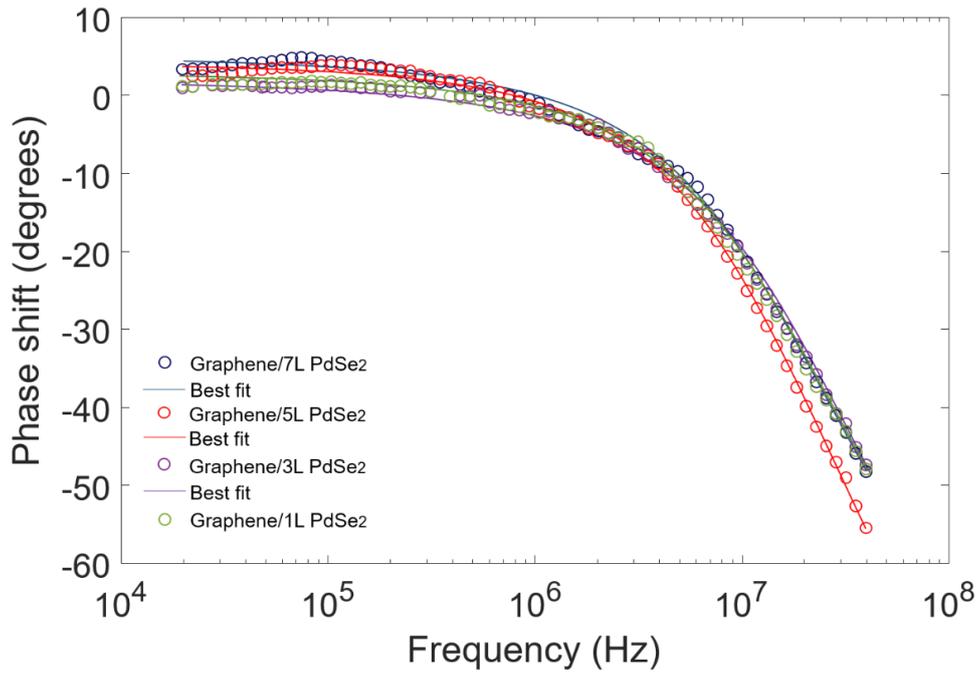

**Fig. S6.** FDTR data sets measured in graphene/PdSe$_2$ films of different thickness grown on silicon carbide and the corresponding best model fits in the whole frequency range (20 kHz-45 MHz).

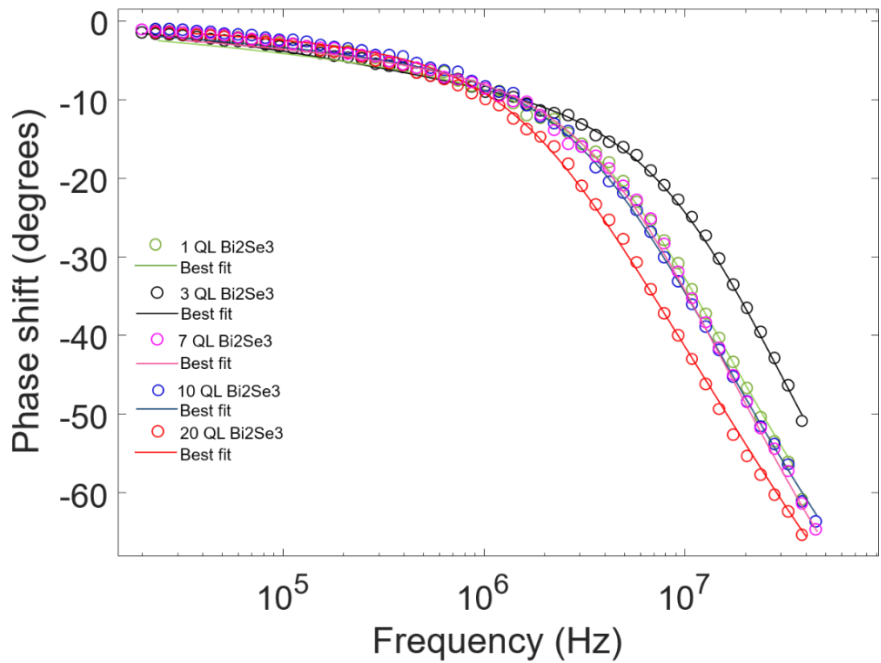

**Fig. S7.** FDTR data sets measured in 1, 3, 7, 10 and 20 QL Bi$_2$Se$_3$ films on sapphire substrate and the corresponding best model fits in the whole frequency range (20 kHz-45 MHz).

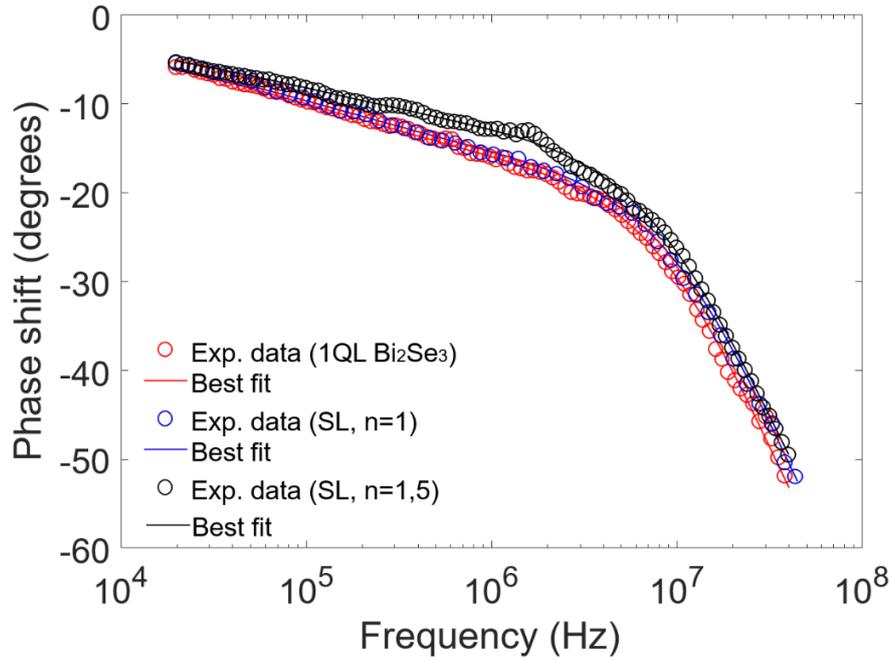

**Fig. S8.** FDTR data measured in 1 QL Bi$_2$Se$_3$ (red circles) and layered Bi$_2$Se$_3$/MoSe$_2$ SLs with periods n = 1, 1.5 and the corresponding best model fits in the whole frequency range (20 kHz-45 MHz).

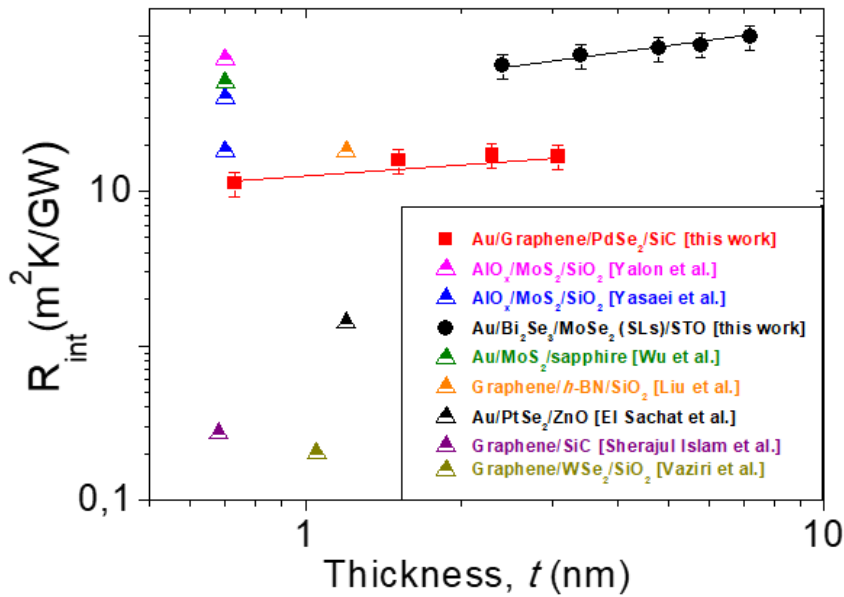

**Fig. S9.** Total interface thermal resistance values, $R_{int}=R_1+R_2$, extracted by the FDTR measurements taking into account the multilayer geometry of Bi$_2$Se$_3$/MoSe$_2$ SLs and graphene/PdSe$_2$ heterostructures. The errors bars estimated by the standard deviation of several

measurements, including the numerical errors from the fits. For comparison, we plot previous interface thermal resistance values measured in 2D/2D and 3D/2D interfaces.[1-7]

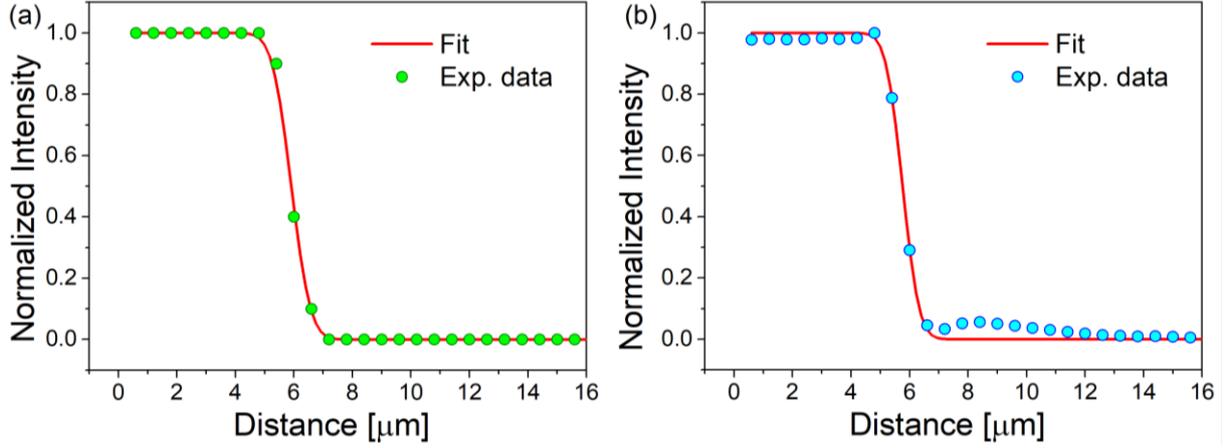

**Fig. S10.** Typical examples of measured (a) pump and (b) probe spot sizes in Au/ $Bi_2Se_3$/$MoSe_2$ (SLs)/STO stacks. The spot size of each measurement was measured by using the knife's edge method. The edge of the Au transducer layer was used as a sharp edge to measure the intensity of the reflected light as a function of stage position. The beam intensity as a function of translation distance was fitted to an error function curve[8] and the $1/e^2$ radius of this curve was taken as the laser spot radius.

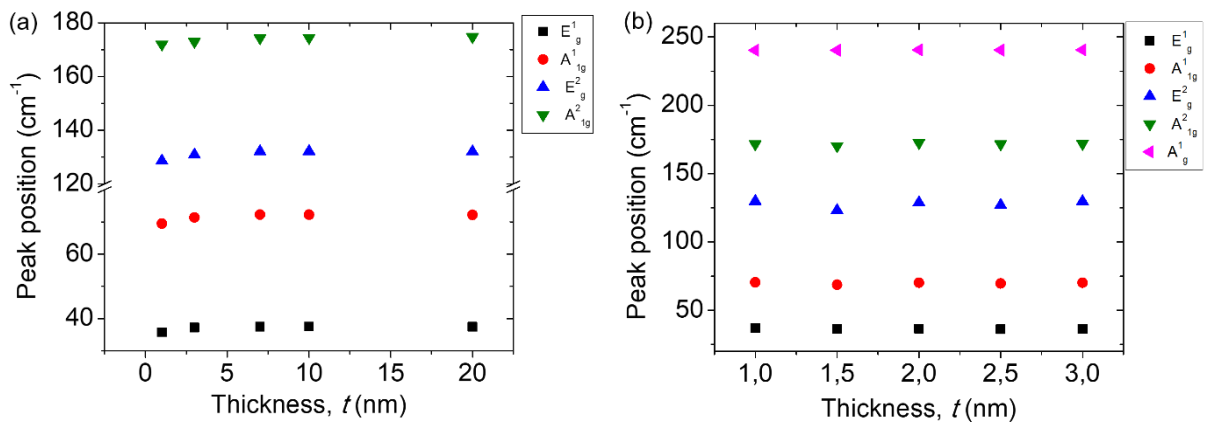

**Fig. S11.** Peak position of all the Raman active modes versus thickness in (a) $Bi_2Se_3$ films and (b) $Bi_2Se_3$/$MoSe_2$ SLs.

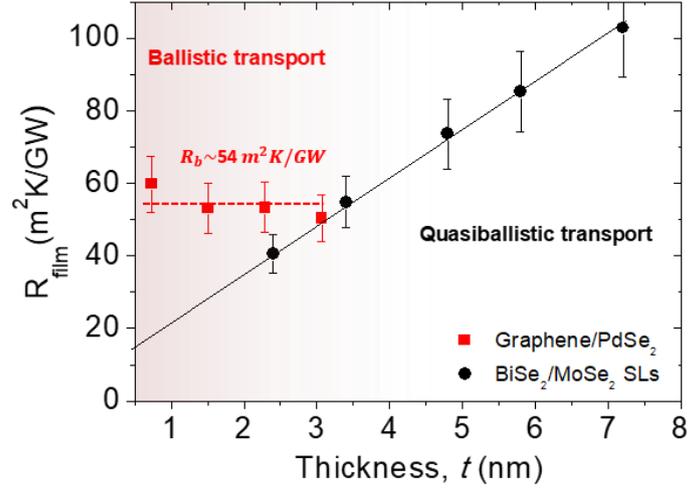

**Fig. S12.** The volumetric cross-plane thermal resistance, $R_{\text{film}} = t/k_z$, as a function of film thickness $t$ for the $Bi_2Se_3/MoSe_2$ SLs (black solid circles) and graphene/PdSe$_2$ heterostructures (red solid squares).

## 3. Tight-binding phonon calculations

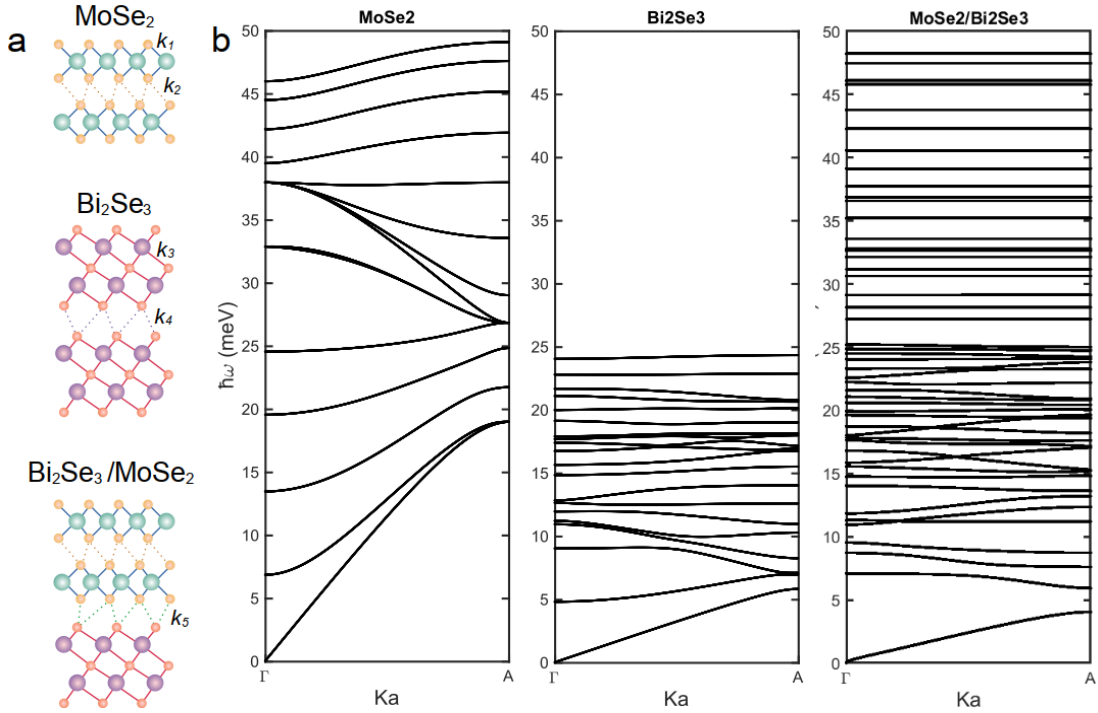

**Fig. S13. Tight-binding phonon band structure.** (a) The schematic structure of layers vdW structures, phonon band structure of (b) MoSe$_2$, (c) Bi$_2$Se$_3$ and (d) Bi$_2$Se$_3$/MoSe$_2$ hetero structure.

**Table S1.** The parameters used for to construct the tight binding model with respect to the reference energy $t$.

| Force constant | $k_1$ | $k_2$ | $k_3$ | $k_4$ | $k_5$ |
|---|---|---|---|---|---|
| | $t$ | $t$ | $t/5$ | $t/6$ | $t/2$ |

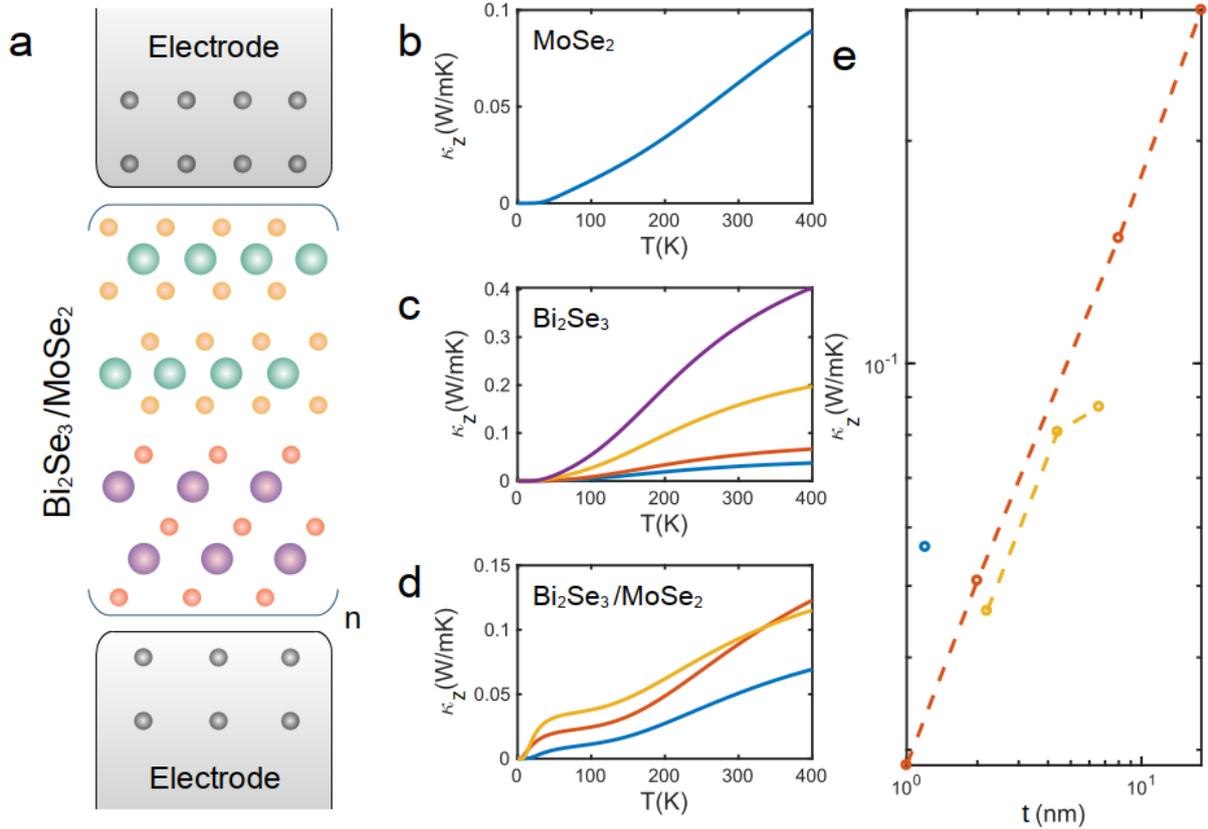

**Fig. S14. Calculated cross-plane thermal conductivity of MoSe$_2$, Bi$_2$Se$_3$ and Bi$_2$Se$_3$/MoSe$_2$ heterostructures considering a weak coupling with the electrodes.** (a) The schematic structure of layers vdW structures between electrodes. The thermal conductivity as a function of temperature of (b) MoSe$_2$, (c) Bi$_2$Se$_3$ and (d) Bi$_2$Se$_3$/MoSe$_2$ heterostructures. (e) The thermal conductivity as a function of thickness for MoSe$_2$, Bi$_2$Se$_3$ and Bi$_2$Se$_3$/MoSe$_2$ heterostructures at room temperature.

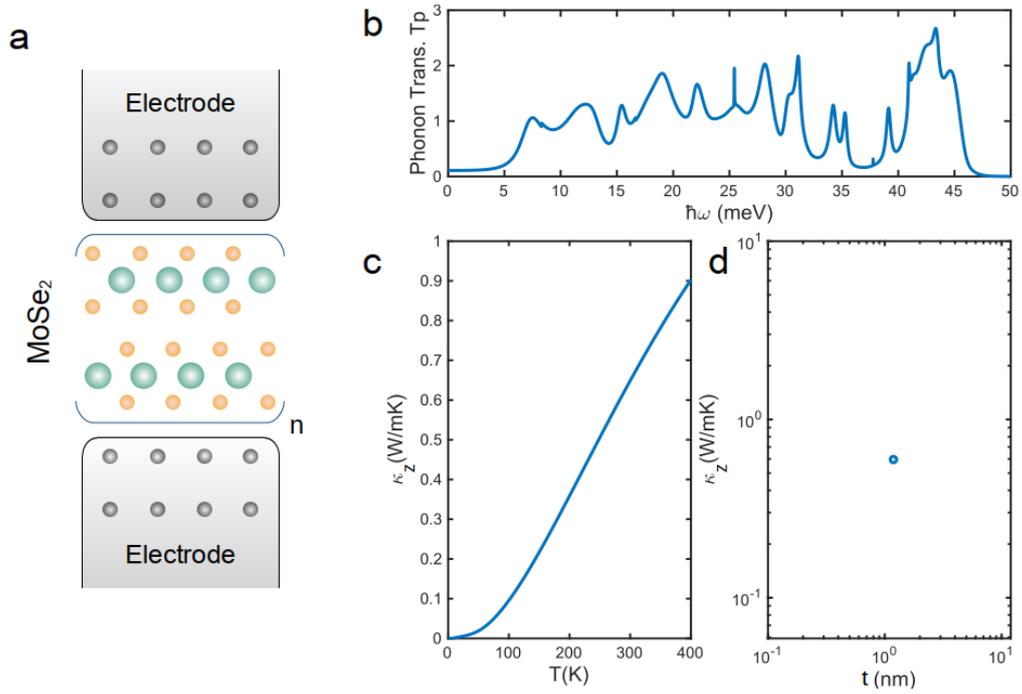

**Fig. S15. Calculated cross-plane thermal conductivity of MoSe$_2$ considering a strong coupling with the electrodes.** (a) The schematic structure of MoSe$_2$ between electrodes, (b) phonon transmission coefficient as a function of phonon energy, (c) thermal conductivity as a function of temperature and (d) thermal conductivity as a function of thickness at room temperature.

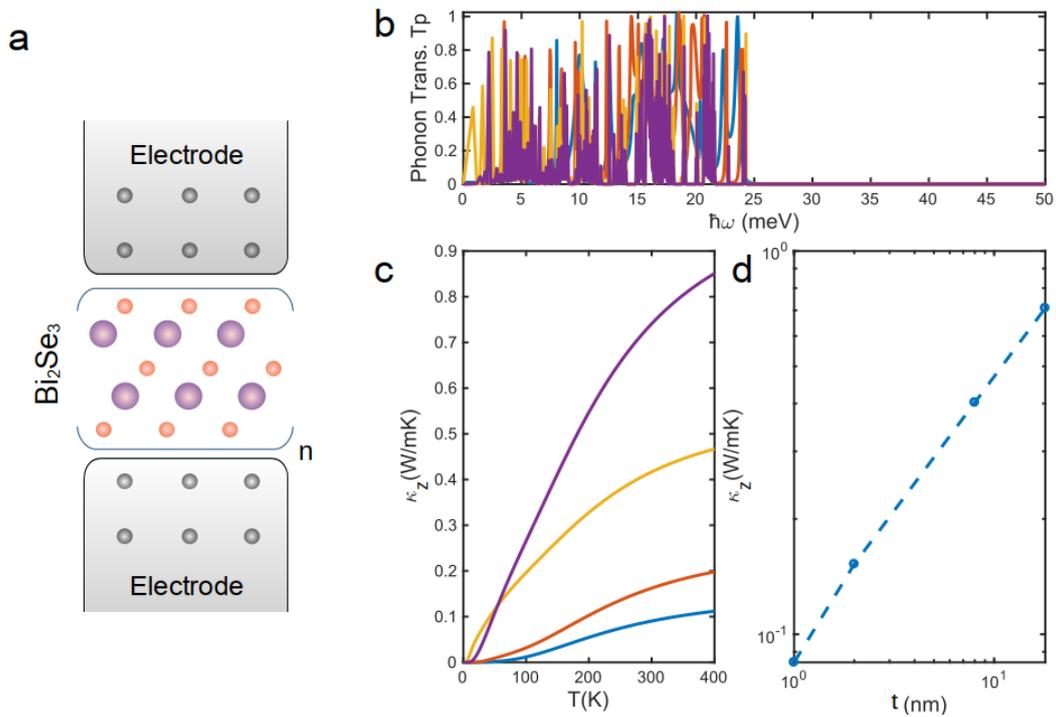

**Fig. S16. Calculated cross-plane thermal conductivity of Bi$_2$Se$_3$ considering a strong coupling with the electrodes.** (a) The schematic structure of Bi$_2$Se$_3$ between electrodes, (b) phonon transmission coefficient as a function of phonon energy, (c) thermal conductivity as a function of temperature and (d) thermal conductivity as a function of thickness at room temperature.

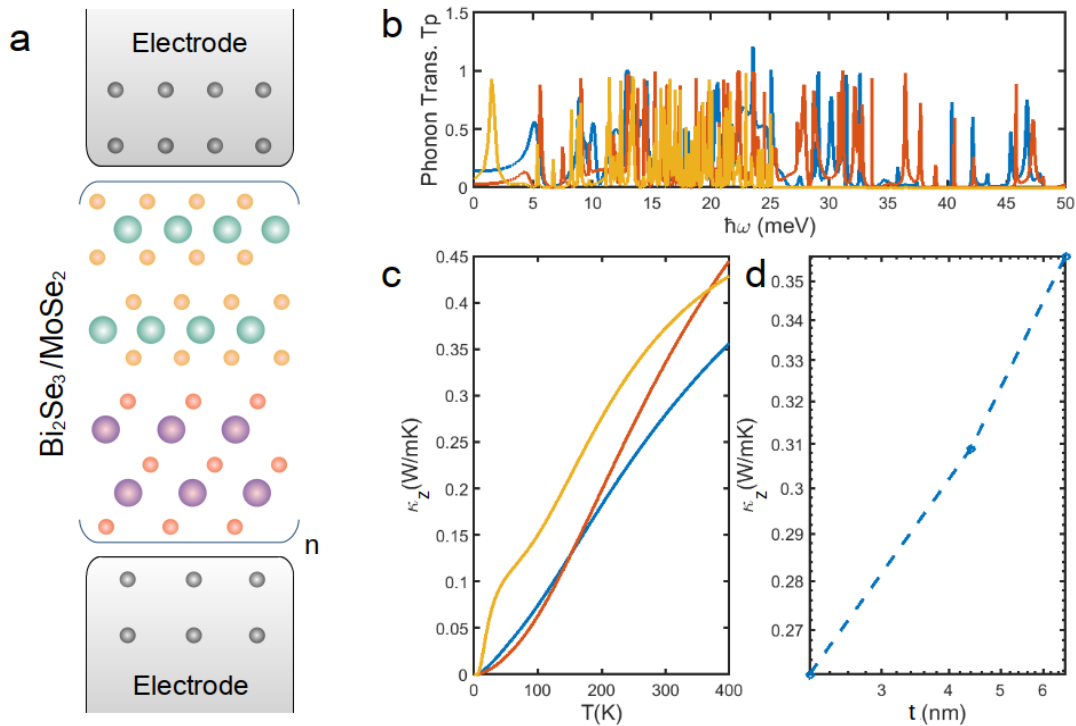

**Fig. S17. Calculated cross-plane thermal conductivity of Bi$_2$Se$_3$/MoSe$_2$ heterostructures considering a strong coupling with the electrodes.** (a) The schematic structure of Bi$_2$Se$_3$/MoSe$_2$ between electrodes, (b) phonon transmission coefficient as a function of phonon energy, (c) thermal conductivity as a function of temperature and (d) thermal conductivity as a function of thickness at room temperature.